\documentclass[12pt,preprint]{aastex}

\def\subsun{\mbox{$_{\normalsize\odot}$}}
\def\kms{\ifmmode {\rm km\,s}^{-1} \else km\,s$^{-1}$\fi}

\def\simlt{\lower.5ex\hbox{$\; \buildrel < \over \sim \;$}}

\begin{document}

\title{{\em Hubble Space Telescope}\ Imaging of the Post-Starburst Quasar UN~J1025$-$0040:  Evidence for Recent Star Formation}
\author{Michael S. Brotherton\altaffilmark{1}, Matthew Grabelsky\altaffilmark{2}, Gabriela Canalizo\altaffilmark{3}, Wil van Breugel\altaffilmark{3}, 
Alexei V. Filippenko\altaffilmark{4}, Scott Croom\altaffilmark{5}, 
Brian Boyle\altaffilmark{5}, Tom Shanks\altaffilmark{6}}

\altaffiltext{1}{Kitt Peak National Observatory, National Optical Astronomy Observatories,
950 North Cherry Avenue, P. O. Box 26732, Tucson, AZ 85726; mbrother@noao.edu.
The National Optical Astronomy Observatories are operated by the Association of Universities for Research in Astronomy, Inc., under cooperative agreement with the National Science Foundation.}
\altaffiltext{2}{Department of Physics and Astronomy, Rice University, Houston, TX; grabelsky@hotmail.com}
\altaffiltext{3}{Institute of Geophysics and Planetary Physics, Lawrence Livermore National Laboratory, 7000 East Avenue, P.O. Box 808, L413, Livermore, CA 94550}
\altaffiltext{4}{Department of Astronomy, 601 Campbell Hall, University of California, Berkeley, CA 94720-3411}
\altaffiltext{5}{Anglo-Australian Observatory, PO Box 296, Epping, NSW 2121, Australia}
\altaffiltext{6}{Department of Physics, University of Durham, South Road, Durham DH1 3LE}
\begin{abstract}

We present new {\em Hubble Space Telescope} ({\em HST}) WFPC2 
images of the post-starburst quasar UN~J1025$-$0040, which contains 
both an active galactic nucleus (AGN) and a 400-Myr-old nuclear starburst 
of similar bolometric luminosity ($\sim10^{11.6} L\subsun$).  
The F450W and F814W images resolve the AGN from the starburst and show that the bulk 
of the star light ($\sim$ 6 $\times$ 10$^{10}$ M$_\sun$) is contained within a 
central radius of about 600 parsecs, and lacks clear morphological structures
at this scale.  Equating the point-source light in each image with the AGN contribution,
we determined the ratio of AGN-to-starburst light.  This ratio is 69\% 
in the red F814W image, consistent with our previous spectral analysis, 
but $\leq 50\%$\ in the blue F450W image whereas we had predicted 76\%.  
The {\em HST} images are consistent with previous photometry, ruling out 
variability (a fading AGN) as a cause for this result.  We can explain 
the new data if there is a previously unknown young stellar population 
present, 40 Myr or younger, with as much as 10\% of the mass of 
the dominant 400-Myr-old population.  This younger starburst may represent
the trigger for the current nuclear activity.  The multiple starburst ages
seen in UN~J1025$-$0040 and its companion galaxy indicate a complex 
interaction and star-formation history.

\end{abstract}

\section{Introduction}

Exciting progress has been made bolstering our understanding of
the connection between active galactic nuclei (AGNs) and starbursts.
It appears that essentially every large galaxy harbors a supermassive dark
object (presumably a black hole) some 0.15\% of the mass of its spheroidal
component (Magorrian et al. 1998; Gebhardt et al. 2000a), although the black
hole mass perhaps correlates more tightly with the velocity dispersion of the 
host bulge (e.g., Gebhardt et al. 2000b; Ferrarese \& Merritt 2000a).
These are plausibly ``dead'' quasars, which once burned while undergoing
high accretion episodes.  Furthermore, black hole masses for ``live'' quasars
determined from reverberation techniques also correlate with the bulge mass
and velocity dispersion (Gebhardt et al. 2000b; Ferrarese et al. 2001).
Consistent with these results, there exists an upper limit to the luminosity 
of AGNs that depends on their host galaxy stellar mass (e.g., 
McLeod, Rieke, \& Storrie-Lombardi 1999).  Finally, there is the global 
similarity in the strong evolution of the quasar luminosity function and 
the star-formation rate (e.g., Boyle \& Terlevich 1998).

These points have sparked much recent theoretical interest in hierarchical
galaxy formation and related scenarios featuring direct relationships
between star formation and the growth/fueling of central black holes
(e.g., Burket \& Silk 2001, and references therein).
AGN activity, in fact, may reflect a fundamental stage of galaxy evolution.

The discovery of a large population of
ultraluminous infrared galaxies (ULIRGs; $L_{IR}$ $> 10^{12} L\subsun$) that
are strongly interacting merger systems heated by both starburst and AGN power sources 
led Sanders et al. (1988) to hypothesize that ULIRGs represent the initial 
dust-enshrouded stage of a quasar (see also Sanders \& Mirabel 1996 for a review).  
The evidence for this evolutionary scenario is circumstantial, but the
hypothesis is supported by {\em HST} images of warm ULIRGs and optical
QSOs that show similarities such as knots of recent star formation
(e.g., Surace et al. 1998; Boyce et al. 1996; Bahcall et al. 1997).
But Boyce et al. (1996) suggested that the $IRAS$-selected quasars are the
products of {\em young} spiral mergers, not old mergers. Recent
$ISO$\ results may also be incompatible with a strict evolutionary scenario:
Genzel et al. (1998) surveyed ULIRGs and found that half contain simultaneously
an AGN and starburst activity in a 1-2 kpc circumnuclear ring (these
very young star-forming regions have ages 10-100 Myrs), but that the presence of
an AGN is {\em not} correlated with the merger age (see also
Canalizo \& Stockton 2001).

During the course of identifying quasar candidates in the AAT-2dF quasar
survey (Croom et al. 2001) coincident with radio sources in the NRAO
VLA Sky Survey (NVSS; Condon et al. 1998), we discovered UN J1025$-$0040
(Brotherton et al. 1999a). This object's unusual spectrum
shows the broad Mg II emission line of a quasar as well as the large Balmer jump
and deep Balmer absorption lines reminiscent of an A-type star,
all at $z=0.634$. The spectrum may be categorized as a ``Q + A"
(quasar plus post-starburst) spectrum similar to those of post-starburst
galaxies which have ``k + A" spectra due to the combination of young and old
stellar populations.  We modeled the stellar content of UN J1025$-$0040 using
stellar synthesis population models (Bruzual \& Charlot 1996), and found that
a 400-Myr-old instantaneous starburst provides an excellent match to the
Balmer jump and stellar absorption lines.  In conjunction with the broad emission
lines and power-law spectrum of a blue quasar, the observed spectrum could be 
completely accounted for, although not uniquely.  A population of very young stars
could also be present, but such a population lacks strong features
in the observed spectral region to betray its presence (ultraviolet spectroscopy
or imaging analysis, as in this paper, is needed).

The estimate of the initial mass of the 400-Myr-old starburst was nearly 
$10^{11}$ $M_{\sun}$, or the mass of our own Milky Way Galaxy.  Subtracting off 
this model left a residual spectrum that resembles a normal if rather excessively 
blue quasar.  The total energy budget of both components appeared similar, 
$10^{11.6}$ $L_{\sun}$. Running the stellar synthesis population model in 
reverse, when the starburst was only 50 Myr old it would have had a 
bolometric luminosity of $10^{12.4}$ $L_{\sun}$ and qualified as an ULIRG.

Deep Keck K$_s$-band imaging with 0.5\arcsec\ seeing failed to
resolve quasar and stellar components although it did reveal a companion 
galaxy 4.2\arcsec\ away that turned out to have a dominant 800-Myr-old stellar
population (Canalizo et al. 2000).  Faint [O II] $\lambda$3727 knots are also
present between UN J1025$-$0040 and the companion.
Certainly this is an interacting system, as are essentially all ULIRGs.

With these results in hand, we were awarded {\em HST} cycle 9
time to image UN J1025$-$0040 with WFPC2 and attempt to resolve the starburst
from the quasar.  The sub-pixel dithered resolution of the WFPC2 PC achieves a
spatial resolution of about 600 parsecs FWHM at $z=0.634$
($H_0=75$ \kms Mpc$^{-1}$, $q_0=0$), and we anticipated seeing a 
$\sim$kpc-scale starburst ring or a double nucleus.  

We do not resolve any such structures (\S\ 2).  We marginally resolve
extended emission (stars) around a point source (the AGN) and see a 
disturbed morphology suggestive of a nuclear merger remnant (\S\ 3).  The ratio
of the point-to-extended light differs from the spectral analysis of Brotherton
et al. (1999a) and indicates the presence of a previously unsuspected
young or on-going starburst (\S\ 4).   We discuss the implications of 
these results (\S\ 6) and summarize our conclusions (\S\ 7).

\section{{\em HST} Images}

We imaged UN J1025$-$0040 in two broad bands, F814W in the red longward of
the redshifted Balmer jump, and F450W in the blue shortward of the 
redshifted Balmer jump.  UN J1025$-$0040 was centered on the Planetary Camera 
(PC) detector of WFPC2 (0.046 arcsec per pixel), and multiple images at each 
band were taken using a sub-pixel dithered box pattern for maximum resolution.
Eight 500 sec observations were made on 13 Apr 2001 (UT) through the
F814W filter, and sixteen 500 sec observations on 20 Apr 2001 (UT) through 
the F450W filter.

We began with the pipeline-calibrated images which already had undergone
bias subtraction and flat fielding.  We combined our dithered images using two
different methods, the DRIZZLE algorithm of Fruchter et al. (1997), and
the algorithm of Lauer (1999).  Both algorithms produce images with twice
the original number of pixels along each axis, 0.023 arcsec per pixel in the
case of the PC.  These combination procedures also efficiently clean cosmic
rays.  An examination of the profiles of the resulting images showed 
that the algorithm of Lauer (1999) produced sharper images.  For 
the blue F450W images, the FWHMs of cross-cuts through the 
point-source-dominated peaks were 3.165 pixels versus 3.875 pixels,
and for the red F814W image the FWHMs were 4.537 and 5.042 pixels, respectively,
with the Lauer reduction superior in each case.
We therefore used the images produced by the Lauer (1999) algorithm 
for our analyses, which are shown in Figures 1 and 2.

The compact companion galaxy 4.2\arcsec\ to the south-southwest is detected and
resolved for the the first time in these {\em HST} images.  
In the F450W image the companion spans a total spatial extent of 0.23$\arcsec$\ 
and in the F814W image 0.35$\arcsec$; the size in the red image is then
2 kpc for $h = 0.75$, which was approximately the limit reported by
Canalizo et al. (2000).  Table 1 gives the photometry of the integrated flux of
UN J1025$-$0040 and its companion through the F450W and F814W filters.
Figure 3 shows that the photometry is in good agreement with that reported by 
Canalizo et al. (2000), indicating that the AGN has not varied significantly
between the two epochs.  This point will be important in reaching some of 
our conclusions below.  The magnitudes of the companion galaxy are also
entirely consistent with our previous ground-based photometry.

\section{Image Analysis}

From the fully reduced images it is apparent that we have failed to clearly
resolve any significant starburst structures (e.g., a second nucleus or 
a ring).  The images do resolve much of the centrally concentrated 
starburst, which is extended over a diameter of $\sim$1 kiloparsec around the sharper 
point source nucleus that is the AGN.  Some spurs of low surface brightness
extended outward, the most notable being toward the northeast.  A low
surface brightness plateau extends to the west about 0.5 arcsec.
The outermost surface brightness contour in the red image shows is perhaps
suggestive of a merger.

There were no suitable stars in the image field of the PC, so we used 
the Tiny Tim software (Krist \& Hook 1999) to produce model PSFs in conjunction with
the spectral shape of the point source, a blue quasar, in this case using a composite 
quasar spectrum (Brotherton et al. 2001).  We tried different values of the jitter 
parameter to find the best estimate for the PSF in each image, as telescope breathing
and detector effects make the PSF impossible to model perfectly.  The jitter added was
14 mas for the F450W model, and 18 mas for the F814W model.

Once we had good estimates of the PSFs, we used the Richard-Lucy algorithm 
to deconvolve our images.  These deconvolved images did not reveal any
significant additional detail.  We have a disturbed nuclear region with a 
centrally concentrated, amorphous structure marginally resolved at the PC scale.
The very low level extended (arcsecond scale) starlight is energetically negligible
compared to the centrally concentrated light.  This surrounding ``fuzz''
has colors that are very red, $m(F450W)-m(F814W)=2.0$\ mag in the northeast and
$m(F450W)-m(F814W)=1.6$\ mag in the west, indicating an old (1+ Gyr) stellar population.

We also created PSF-subtracted images to determine the ratio of point source
to extended flux.  The minimum amount of point source present was estimated
by subtracting off increasing amounts of the PSF model until a smooth residual 
was obtained that did not have a central dip.  The maximum amount was estimated 
by subtracting off increasing amounts of the PSF model until zero flux was reached
in the residual.  This uncertainty dominates those present in the choice of PSF
model.  We estimate that the extended emission represents 
$52\pm4\%$\ of the total flux in the F450W image, and $69^{+4}_{-5}\%$\ in the
F814W image. 

\section{Spectral Analyses: Quasar and Starbursts}

Figure 4 shows an adaptation of a figure from Brotherton et al. (1999a),
in which a 400-Myr-old instantaneous starburst accounts for the stellar
light in UN J1025$-$0040 and a rather blue AGN accounts for the remainder.
This previous analysis indicated that the starburst contributes some
two-thirds of the red light, and one quarter to one third of the blue light.
The size of the Balmer jump and the depths and equivalent width ratios of
the stellar absorption lines provided excellent constraints in the 
red part of the spectrum, but there are few stellar features in the blue
to check the model.

The image analysis described in the previous section provides an independent 
estimate of the fractions of AGN and starburst in each band.  While there
is good agreement in the red F814W band between the starburst fractions
determined by the image analysis and the previous spectral analysis,
the blue F450W band shows a surprise.  The starburst fraction in the blue
is {\em half} the total, not $\sim$30\% as expected, a significant difference.

We have an excess of extended blue light, so it cannot arise from direct AGN
light.  The polarization is low (1-2\%; Brotherton et al. 1999a) so it cannot 
arise from indirect, scattered light from the AGN, and the size of the
scattering region would have to be implausibly large.

Single-age starbursts cannot simultaneously explain the spectral features
in the red part of the spectrum and the blue-red spectral energy 
distribution, even when allowing the several free parameters to vary
(e.g., metallicity, dust, initial mass function).

Very young stars can contribute significant blue light, while at the same
time having only weak Balmer jumps and weak absorption lines.  Such a 
population, added to a 400-Myr-old population, can make a composite 
starburst spectrum bluer yet allow it to retain essentially the same
Balmer jump and absorption line equivalent width
ratios (which are unaffected by the diluting quasar light).
Figure 5 illustrates again the spectrum of 
UN~J1025$-$0040 as in Figure 4, but also shows the total light photometry, 
the estimated starburst photometry, and a composite starburst population
model.  The composite model was created using Bruzual \& Charlot (1996) 
stellar synthesis population models with two instantaneous starbursts
(Salpeter IMF, solar metallicity, no dust), one 400 Myr old, the other 40 Myr 
old or younger (10 Myr old in the case of Fig. 5), scaled to match the 
combination of imaging and spectral constraints.  Young starbursts older
than 40 Myr are increasingly red, have larger Balmer jumps, and more
pronounced absorption lines; such populations are still bluer
than the 400 Myr old model and capable of meeting the imaging constraints
when added in the correct proportions, but will fail to simultaneously
match the spectral constraints of the Balmer jump and absorption lines.

As discussed previously by Brotherton et al. (1999a), varying  
the model parameters (primarily metallicity) can result in somewhat 
different ages.  In the face of increasing complexity of this system and 
the limited data so far available, determining the exact parameters and 
ages is not yet possible.  Nevertheless, we have a robust result that a 
significantly younger stellar population is present in the nucleus of 
UN J1025$-$0040.

The residual quasar spectrum implied by this analysis is not as blue
as originally suggested by Brotherton et al. (1999a).  The quasar continuum
can be approximated by a power-law continuum where the optical spectral
index $\alpha = -0.5$, where $f_{\nu} \propto \nu^{\alpha}$.  This is
a very typical optical power-law index (e.g., Francis 1993).

How young (and hence also how massive) is the young starburst?
How does its presence affect our previous conclusions regarding
the older, more massive starburst?  Figure 6 shows a range of viable
models.  All models shown are combinations of dust-free instantaneous 
starbursts employing Salpeter initial mass functions and solar metallicities,
and the composite quasar spectrum of Brotherton et al. (2001) based
on the FIRST Bright Quasar Survey (FBQS; White et al. 2001).  The combination
of bright optical selection and deep radio selection in the FBQS is
rather similar to that used for the 2dF-NVSS selection that identified
UN J1025$-$0040.  This composite quasar spectrum has a power-law index
of $\alpha = -0.46$, consistent with that determined above.

The scaling, and hence mass, of the 400-Myr-old starburst is rather
insensitive to the age of the younger population.  For the updated 
photometry here and for $H_0=75$ \kms Mpc$^{-1}$, $q_0=0$, the oldest 
composite starburst model of Figure 6 contains a 400-Myr-old starburst with 
$5.7\times10^{10} M\subsun$, plus a 40-Myr-old starburst with 
$6.7\times10^{9} M\subsun$,
or 10\% of the mass of the older starburst.  For the youngest models of
Figure 6 (0 and 1.5-Myr-old instantaneous bursts), the starburst masses are 
down another order of magnitude, or about 1\% of the mass of the older starburst.
At a good fraction of 10$^9 M\subsun$, this still qualifies as a very massive
starburst in its own right.  The exact parameters are
not well determined and require additional constraints in the form of 
ultraviolet spectroscopy.  The caveats of Brotherton et al. (1999a) regarding
the uncertainties on the age and mass of the older starburst should also 
be kept in mind.

\section{Discussion}

The primary aim of obtaining the {\em HST} images was to resolve the starburst
from the AGN and identify any morphological structures that might provide a
guide as to the origin of such an extremely massive starburst.  
In that, we have been partially successful, insofar as we can distinguish a 
point source from a more diffuse region of centrally concentrated
starlight.  Since the ratio of starlight to quasar in the red part of 
spectrum is just as predicted from the spectral analysis of Brotherton 
et al. (1999a), we are probably not missing any core of starlight still 
unresolved from the AGN.

We were not successful in identifying any clearly recognizable star-forming
structures, such as a starburst ring as has been seen in some other AGNs
(e.g., NGC~985, Rodriguez Espinosa \& Stanga 1990).  At these redshifts the {\em HST} 
resolution is only probing the marginally sub-kiloparsec scale, so it is still 
possible that such an organized structure exists on smaller scales but is not yet 
resolvable as such.

These shallow images do not reveal as much detail outside the nucleus as do the 
deeper low-resolution ground-based images (Brotherton et al. 1999a; 
Canalizo et al. 2000).  The nuclear structure is revealed to be non-uniform and
suggestive of an evolved merger product, perhaps similar to Arp 220 
(e.g., Scoville et al. 1998), but not as dusty. 
Whether or not UN J1025$-$0040 is an actual merger product or merely disturbed and
interacting will likely require much deeper high-spatial-resolution images.
Interpreting UN J1025$-$0040 as a merger product, however, would be entirely 
consistent with the evolutionary hypothesis of Sanders et al. (1988)
in which the end result of massive mergers would be such a post-starburst quasar.
This evolutionary hypothesis has been recently boosted by detailed work on
the IRAS 1-Jy sample of ULIRGs (e.g., Veilleux, Kim, \& Sanders 1999) which shows 
an increasing fraction of luminous AGNs with increasing merger age.  
This is a large, complete sample that includes primarily the most luminous ULIRGs
for which the evolutionary hypothesis may hold true overall; contrary results,
like those of Genzel et al. (1998), may result from small samples including
less luminous ULIRGs.  This is an ongoing area of debate.

The new result of the presence of a young ($\leq$40 Myr) starburst adds to this
debate, indicating that specific systems can indeed be complex.  
Canalizo \& Stockton (1997) suggested that in an evolutionary progression, the age 
of a starburst is a clock indicating the age of the AGN activity (Canalizo \& Stockton 2001 
strengthens the original arguments but also provides additional caveats).  
This assumes that the fueling of the nuclear
starburst and the central black hole occurs at the same time, and that the
black hole continues to be fueled.  Determining the ages of starbursts in 
AGN systems would then permit the quasar lifetime to be estimated, and the
quasar duty cycle as well.  These numbers would complement those determined from
combining the evolution of the quasar luminosity function and the population of relic 
black holes today in the local universe.  The complication of our new result
is that there has been a substantial starburst in UN J1025$-$0040 much more
recently than 400 Myr, and this starburst may represent the appropriate clock
for the current activity.  The AGN may well not shine steadily, but undergo 
bursts that correspond to events that provide fuel and also incite
star formation.

Our measurements of the starburst masses in the nucleus of UN J1025$-$0040, 
a total of 6-7 $\times$ 10$^{10}$ M$_\sun$, represent an estimate or at least lower 
limit to the host galaxy bulge mass.  We can estimate the AGN properties,
black hole mass and fraction of the Eddington luminosity at which it shines,
using techniques alluded to in the introduction (following the formalism
and choice of parameters of Merritt \& Ferrarese 2001, which includes
our chosen cosmology).  A black hole mass is estimated by assuming 
Keplerian motions of ionized gas in a spherical broad emission line region (BLR), 
with a velocity $v_{BLR} = \sqrt{3}/2\ {\rm FWHM} (H\beta)$ at a distance
$R_{BLR} = 32.9(\lambda L_{5100}/10^{44} {\rm ergs\ s^{-1}})^{0.7}$ light
days (Kaspi et al. 2000).  The FWHM of the H$\beta$ line is rather
uncertain given the contamination from the strong stellar spectrum.
The FWHM of Mg II is generally consistent with that of H$\beta$,
certainly so within the uncertainty of this sort of black hole estimate,
so we adopt FWHM $(H\beta)$ = 5500 \kms (Brotherton et al. 1999a).  
This yields M$_{BH}$ = $2.7 \times 10^8$ M$\subsun$, and 
M$_{BH}$/M$_{Bulge}$ = 0.04\%, which is within the sample range of local
AGN but smaller than the average $<$M$_{BH}$/M$_{bulge}> \sim$ 0.1\% 
(Merritt \& Ferrarese 2001).  We estimate the bolometric luminosity as 
$L_{bol} = 9 \lambda L_{\lambda}$, where $\lambda L_{\lambda}$ is evaluated at 
5500 \AA\ (Kaspi et al. 2000), obtaining $5.7 \times 10^{11} L\subsun$,
consistent with our previous value (Brotherton et al. 1999a).
For our black hole mass estimate, the Eddington luminosity is 
$1.0 \times 10^{13} L\subsun$, and the Eddington fraction $L/L_{Edd} = 0.05$. 
These values are all consistent with those of other QSOs and Seyfert
galaxies so far studied, so nothing stands out as particularly extreme
other than the starburst.

While the extremely massive starburst of UN J1025$-$0040 might suggest that it is  
an isolated case of limited applicability, it actually represents the prototype 
of an entire {\em class} of post-starburst QSOs.  The 2dF QSO Redshift Survey now 
contains close to 23000 QSOs at $z < 3$\ and $B < 21$ (e.g., Croom et al. 2001).  
Some $\sim$3\% of $z<0.75$ quasars (94/2939) show evidence of Balmer jumps and 
high-order Balmer absorption lines at the 3 $\sigma$ level (Croom et al. in prep.).
Determining the starburst ages and masses in conjunction with the AGN properties
(e.g., black hole mass, Eddington fraction as above) for a large sample
may lead to a consistent picture for the relationship between AGNs and the 
evolution of the galaxies that host them, something that we cannot yet do given 
only the singular case of UN J1025$-$0040.  For instance, we might expect to find 
that the starburst age correlates with M$_{BH}$/M$_{bulge}$ since older quasars 
will have had more time to accrete the available fuel, and inversely correlates with
$L/L_{Edd}$ since that fuel must eventually be exhausted.

Quantitatively developing the evolutionary scenario will probably still be 
insufficient to explain all vectors leading to nuclear activity and their duty cycles.
Besides the 97\% of 2dF QSOs that do not show clear spectroscopic evidence for 
(extremely) massive starbursts and which may not represent evolved ULIRGs, there are 
binary quasars.  There are around a dozen binary quasars currently known, and the 
indications are that the nuclei are triggered {\em early} in a galactic merger process 
by tidal forces when the galaxies are some 50-100 kpc apart (Mortlock et al. 1999).
The post-starburst quasar that is FIRST J1643+3156B, the fainter member of
a binary quasar system (Brotherton et al. 1999b), would fit right into the
post-starburst quasar sample of the 2dF.  High-quality follow-up imaging of that
sample may reveal binary quasars like FIRST J1643+3156 in addition to 
evolved starburst mergers like UN J1025$-$0040.  Starburst age would therefore 
appear to be a superior clock of fuel flow compared to merger age,
in characterizing nuclear activity, and should be measured when possible.

Only a decade ago, debate focused upon whether the majority of quasars
had ``fuzz'' around them or not.  {\em HST} and ground-based near-IR imaging then
opened up the study of quasar host galaxies in general.  Today we are 
starting to characterize their stellar populations and may be on the verge
of reconstructing their evolutionary histories.  In the future, studies of
the evolution of galaxies and AGN may be impossible to separate.

\section{Summary}

High-resolution images using WFPC2 on the Hubble Space Telescope show that 
UN~J1025$-$0040 has a disturbed, compact nucleus, but no easily identifiable
morphological structures.  The photometry is consistent with that of Canalizo 
et al. (2000) indicating the AGN has not varied significantly between the 
time the ground-based data was taken and the {\em HST} images.  The F450W and 
F814W images can be decomposed into a point source (the AGN) and residual,
somewhat extended emission (starlight) concentrated within the central several 
hundred parsecs.  The ratio of AGN to starlight in the red F814W image is
consistent with the previous spectral analysis of Brotherton et al. (1999a),
but the ratio in the blue F450W image is much smaller than expected, indicating
the presence of a young stellar population.  The currently available data
constrain the young stars to be anywhere from a $\sim$40-Myr-old population with 
10\% of the mass of the dominant 400 Myr starburst, to a currently active starburst having 
1\% of the mass -- these would still constitute massive starbursts in their
own right.  The older starburst dominates the light at red wavelengths, and the young
starburst dominates the light at UV wavelengths.  This young starburst may
represent the trigger of the current AGN activity.

\acknowledgments

We thank Tod Lauer for providing his image-reduction software.
Matthew Grabelsky acknowledges support from the NOAO/KPNO
Research Experiences for Undergraduates program funded by the
National Science Foundation.  Support for proposal GO-08703 was provided by 
NASA through a grant from the Space Telescope Science Institute, 
which is operated by the Association of Universities for Research in Astronomy,
Inc., under NASA contract NAS 5-26555.  
A.V.F. acknowledges the Guggenheim Foundation for a Fellowship.
A portion of this work has been performed under the auspices of the 
U.S. Department of Energy, National Nuclear Security Administration by the
University of California, Lawrence Livermore National Laboratory under 
Contract W-7405-ENG-48.


\clearpage

\begin{deluxetable}{lcc}
\tablewidth{0pt}
\tablecaption{Photometry\tablenotemark{a}}
\tablehead{
\colhead{Object} & \colhead{F450W} & \colhead{F814W} \\ 
 & (mag) & (mag) }
\startdata
UN J1025$-$0040 & 19.04 & 19.58 \\
Companion & 25.05 & 24.38 \\
\enddata
\tablenotetext{a}{Magnitude zeropoints are used such that Vega's spectrum integrated 
over the bandpass is exactly magnitude zero.}
\end{deluxetable}

\clearpage

\begin{figure}
\plotone{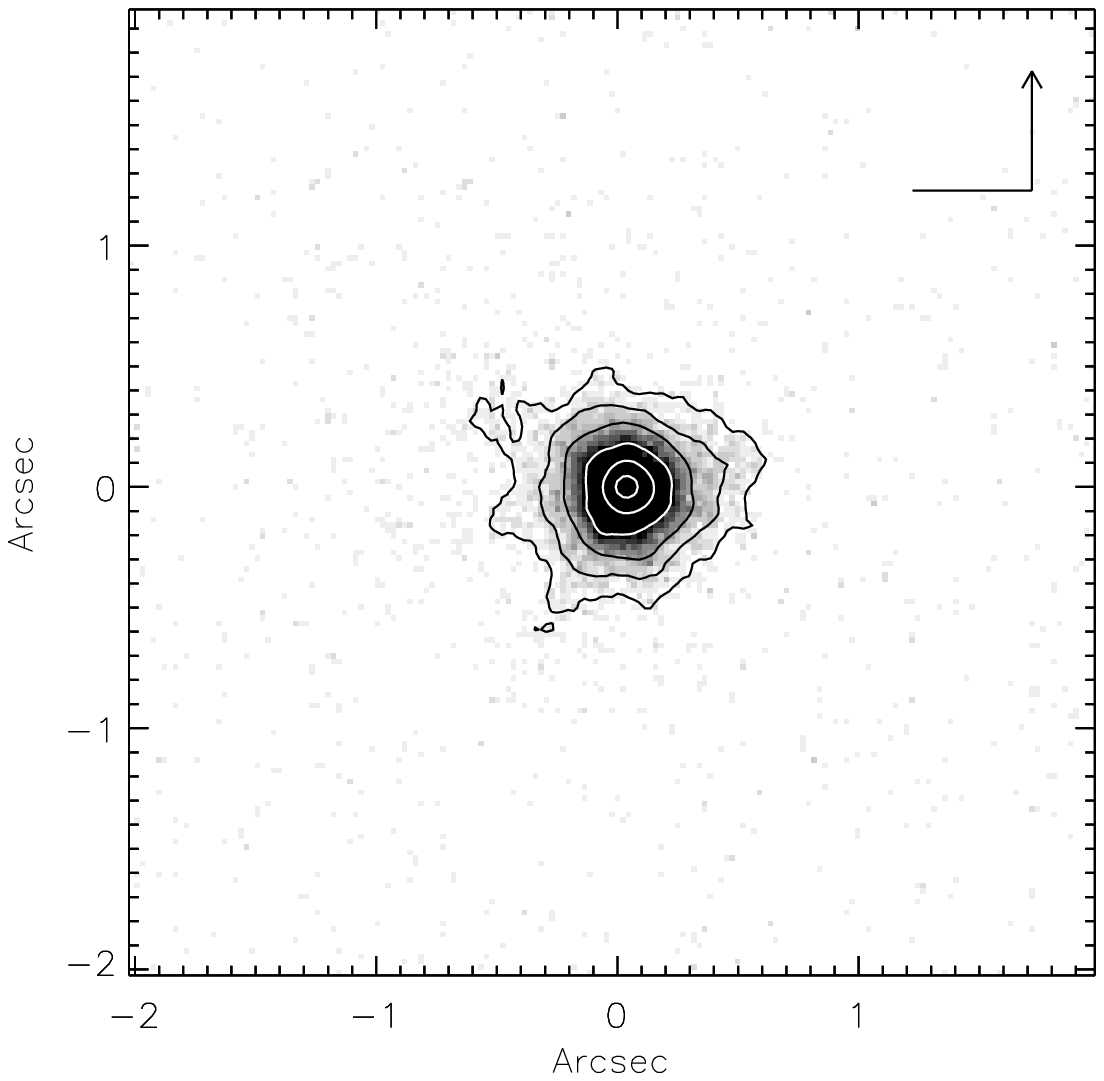}
\caption{{\em HST} image of UN J1025$-$0040, F450W filter on the PC of WFPC2,
using the reduction of Lauer (1999).  The outer contour corresponds
to 21.7 mag per arcsecond$^2$, the inner 15.5 mag per arcsecond$^2$.
North (arrow) and east are indicated.
The scale is 4.3 $h^{-1}$\ kpc/arcsec ($q_o = 0$).\label{fig1}}
\end{figure}

\clearpage

\begin{figure}
\plotone{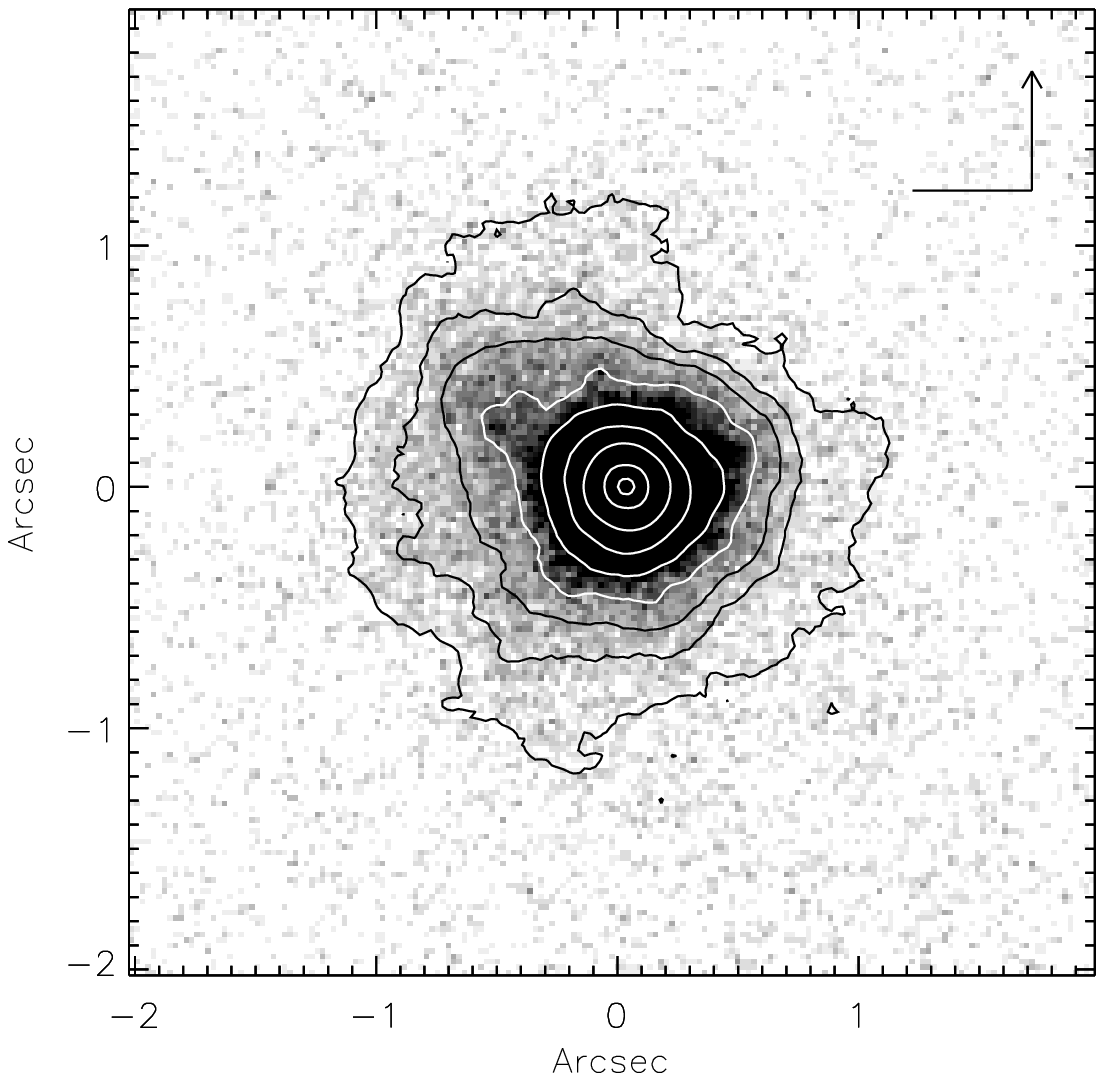}
\caption{{\em HST} image of UN J1025$-$0040, F814W filter on the PC of WFPC2,
using the reduction of Lauer (1999).  The outer contour level corresponds
to 23.4 mag per arcsecond$^2$, the inner 16.8 mag per arcsecond$^2$.
North (arrow) and east are indicated.
The scale is 4.3 $h^{-1}$\ kpc/arcsec ($q_o = 0$).\label{fig2}}
\end{figure}

\clearpage

\begin{figure}
\plotone{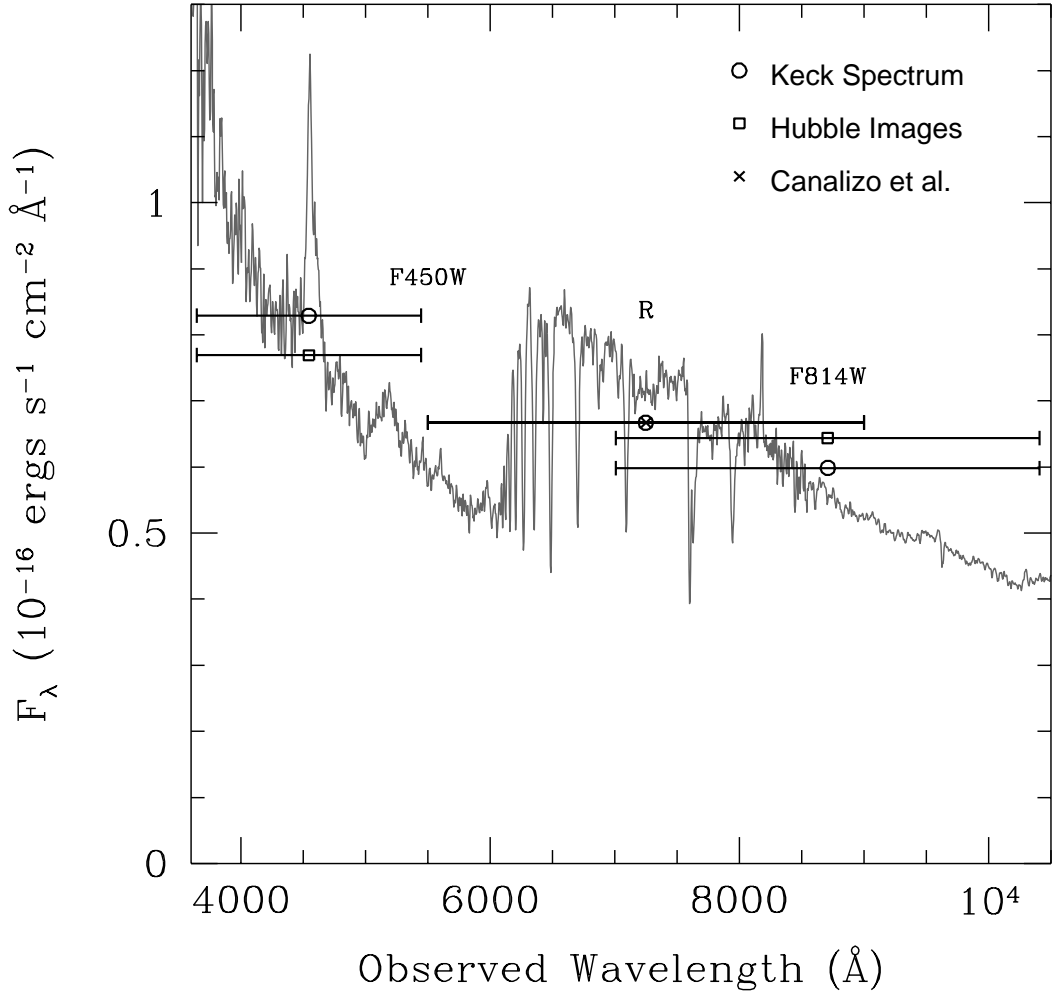}
\caption{Narrow-slit Keck spectrum of UN J1025$-$0040 from Brotherton et al.
(1999a) rescaled to match the $R$-band photometry of Canalizo et al. (2000)
using the IRAF task SBANDS.  The F814W photometry is about 7\% higher,
while the F450W photometry is about 7\% lower.  There is no evidence
for significant AGN variability.\label{fig3}}
\end{figure}

\clearpage

\begin{figure}
\plotone{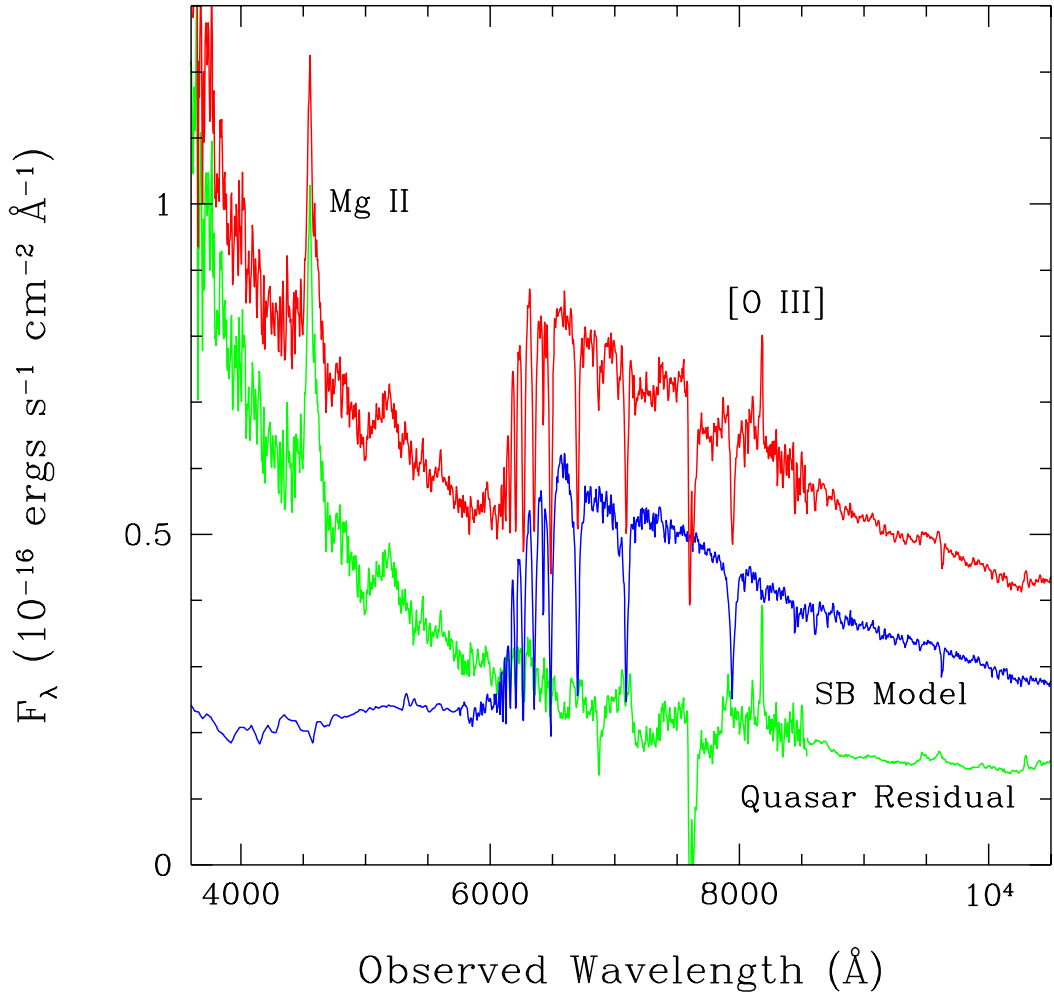}
\caption{Adapted from Brotherton et al. (1999a).  The top red spectrum is that 
of UN J1025$-$0040 (plus a modeled extension at long wavelengths),
a 400 Myr-old instantaneous starburst (blue), and the AGN residual (green) 
that is the difference between the two.\label{fig4}}
\end{figure}

\clearpage

\begin{figure}
\plotone{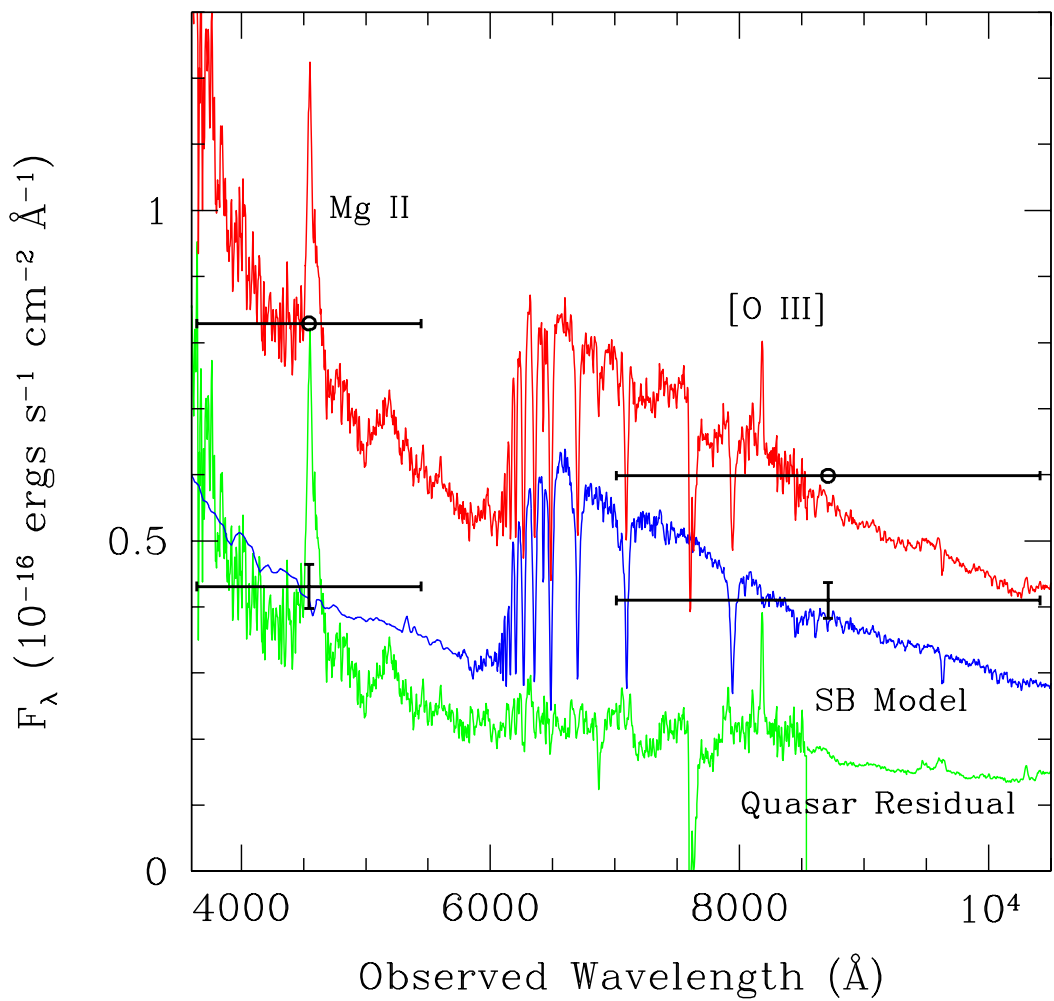}
\caption{We have replotted the rescaled spectrum from Brotherton et al. (1999a)
from Figure 4 (red) with the {\em HST}-band photometry marked (upper points) and the
error bars for the non-PSF (stellar) contribution (lower points).
The new starburst model (blue) is the same 400 Myr-old population as in Figure 4 plus a
young population (in this case 10 Myr) and should be compared to the lower points.  
The quasar residual (green) is the difference between the two spectra above.\label{fig5}
}
\end{figure}

\clearpage

\begin{figure}
\plotone{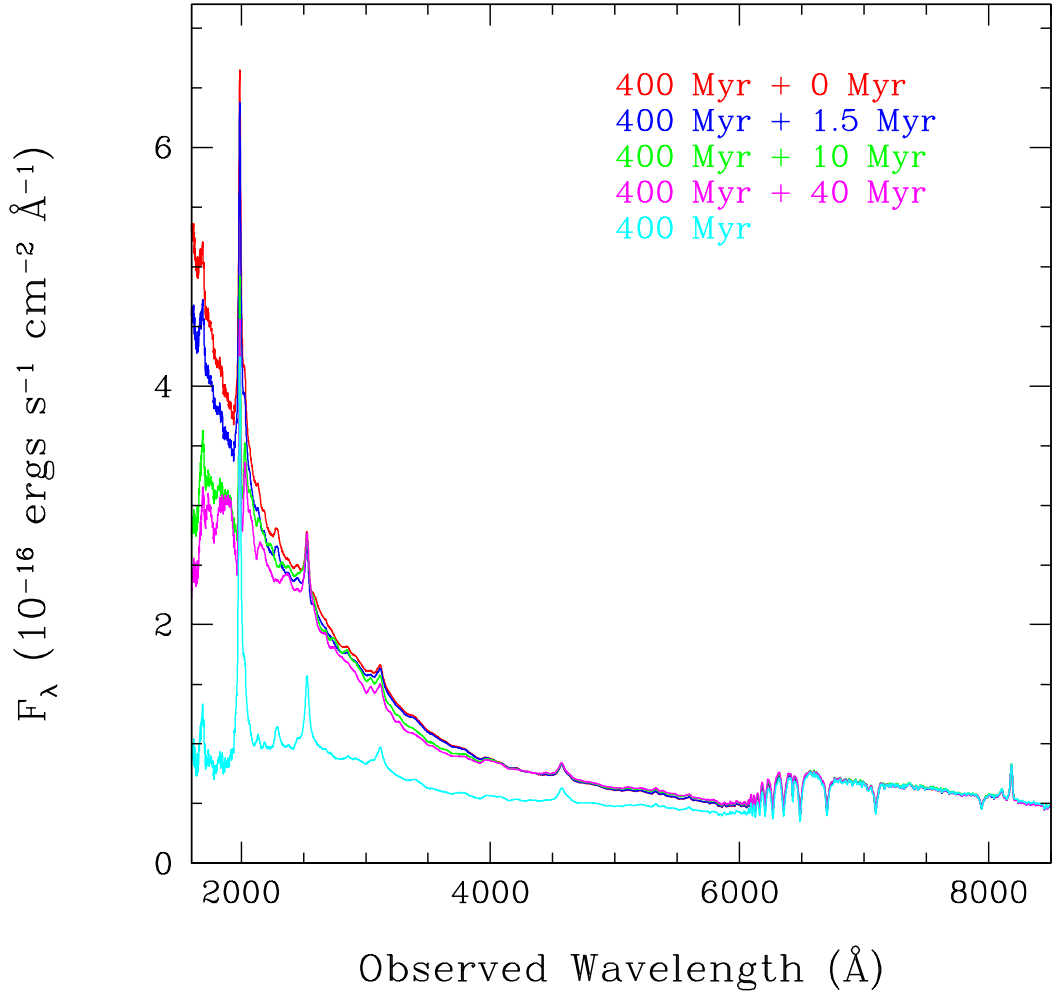}
\caption{A range of viable color-coded models consisting of the composite 
quasar spectrum of Brotherton et al. (2001) and composite starburst models
with ages indicated.  The model with the 400-Myr-old population alone
(light blue) does not match others since the composite quasar spectrum used 
has a typical power-law index and is not sufficiently blue.\label{fig6}
}
\end{figure}

\end{document}